\documentclass[twocolumn,superscriptaddress,prb,amsmath,amssymb,floatfix]{revtex4}

\usepackage{graphicx}
\usepackage{dcolumn}
\usepackage{bm}

\begin{document}


\title{Torque magnetometry studies of  new low temperature metamagnetic
 states in ErNi$_{2}$B$_{2}$C}

\author{D. G. Naugle}
\email{naugle@physics.tamu.edu} \affiliation{Department of Physics,
Texas A\&M University, College Station, TX 77843, USA}

\author{B. I. Belevtsev}
\email{belevtsev@ilt.kharkov.ua} \affiliation{B. Verkin Institute
for Low Temperature Physics and Engineering, National Academy of
Sciences, pr. Lenina 47, Kharkov 61103, Ukraine}

\author{K. D. D. Rathnayaka}
\affiliation{Department of Physics, Texas A\&M University, College Station, TX 77843,
USA}

\author{S.-I. Lee}
\affiliation{National Creative Research Center for
Superconductivity and Department of Physics, Pohang University of
Science and Technology, Pohang 790-784, Republic of Korea}

\author{S. M. Yeo}
\affiliation{National Creative Research Center for
Superconductivity and Department of Physics, Pohang University of
Science and Technology, Pohang 790-784, Republic of Korea}

\begin{abstract}
The metamagnetic transitions in single-crystal ErNi$_2$B$_2$C have
been studied at 1.9 K with a Quantum Design torque magnetometer.
The critical fields of the transitions depend crucially on the
angle between applied field and the easy axis [100]. Torque
measurements have been made while changing angular direction of
the magnetic field (parallel to basal tetragonal $ab$-planes) in a
wide angular range (more than two quadrants). Sequences of
metamagnetic transitions with increasing field are found to be
different for the magnetic field along (or close enough to) the
easy [100] axis from that near the hard [110] axis. The study have
revealed new metamagnetic states in ErNi$_{2}$B$_2$C which were
not apparent in previous longitudinal-magnetization and
neutron studies.
\end{abstract}

\maketitle

The rare-earth nickel borocarbides RNi$_{2}$B$_2$C (where R is a
rare-earth element) show unique superconducting and/or magnetic
properties. In this report, a torque magnetometry study of
metamagnetic transitions at low temperature ($T\approx 1.9$~K) in
single-crystal ErNi$_{2}$B$_2$C is presented. Magnetic states in
this highly anisotropic compound are determined by magnetic
moments of Er ions which lay in the $ab$-planes \cite{zar,lynn}
along the easy axis in the [100] direction. For tetragonal symmetry
inherent in the borocarbides, the [010] direction is an easy axis
as well, so that actually four easy axes are available. The hard
axes in-plane are situated between equivalent easy axes (for example, in
the [110] direction). The crystal lattice of ErNi$_{2}$B$_2$C is
characterized by orthorhombic distortion below 2 K, which,
however, does not have much influence on angular symmetry of
metamagnetic transtions.
\par
In ErNi$_{2}$B$_{2}$C, superconductivity and antiferromagnetism
coexists ($T_c\approx 11$~K and $T_N\approx 6$~K). Below $T_N$ the
magnetic phases are spin-density wave (SDW) states with the
modulation vector $\mathbf{Q}= f \mathbf{a^{*}}$ (or
$\mathbf{b^{*}}$, where $\mathbf{a^{*}}$ and $\mathbf{b^{*}}$ are
reciprocal lattice vectors) \cite{lynn,camp,jensen}. For the AFM
state at zero field $f\approx 0.55$. Ordered moments are
perpendicular to $\mathbf{Q}$. The SDW phase becomes squared-up
below 3 K \cite{zar,lynn}. Below $T_{WF}\approx 2.5$~K a
transition to a weak-ferromagnetic (WF) state occurs, in which a
ferromagnetic moment (about 0.33 $\mu_B$ per Er ion) appears
\cite{canf1}.

\par
With increasing field (applied in the $ab$ plane) several
metamagnetic transitions occur in this compound. The fields of the
transitions depend strongly on the angle $\theta$ between $H$ and a
nearest easy axis (or on the angle $\phi$ between $H$ and a nearest
hard axis) \cite{canf2}. Generally, the ferromagnetic component
increases at each transition reaching the maximum value (about 8
$\mu_B$/Er) at the final transition to the saturated paramagnetic
state at $H\gtrsim 2$~T. The metamagnetic states (except the final
paramagnetic state) remain SDW, only the scalar $f$ of the wave
vector $\mathbf{Q}= f \mathbf{a^{*}}$ changes slightly, but quite
distinctly, at these transitions \cite{camp,jensen}. The known
longitudinal magnetization \cite{canf2,budko} and neutron
diffraction \cite{camp,jensen} studies of metamagnetic transitions
in ErNi$_{2}$B$_{2}$C gave rather different results. The
magnetization studies \cite{canf2,budko} at $T=2$~K  revealed four
metamagnetic transitions for an easy direction (at fields about 0.7
T, 1.1 T, 1.25 T and 2.0 T) and three transitions for a hard
direction (at fields about 1.0 T, 1.4~T and 1.8 T). Also, in the
field range above 2.5 T for field directions not far away from a
hard axis, a weak feature in field dependence of magnetization
was found \cite{canf2} which was interpreted as a final
transition to the paramagnetic state with a critical field that
diverges as $\phi \rightarrow 0$.
\begin{figure}[t]
\vspace{-9pt}
\centering\includegraphics[width=0.9\linewidth]{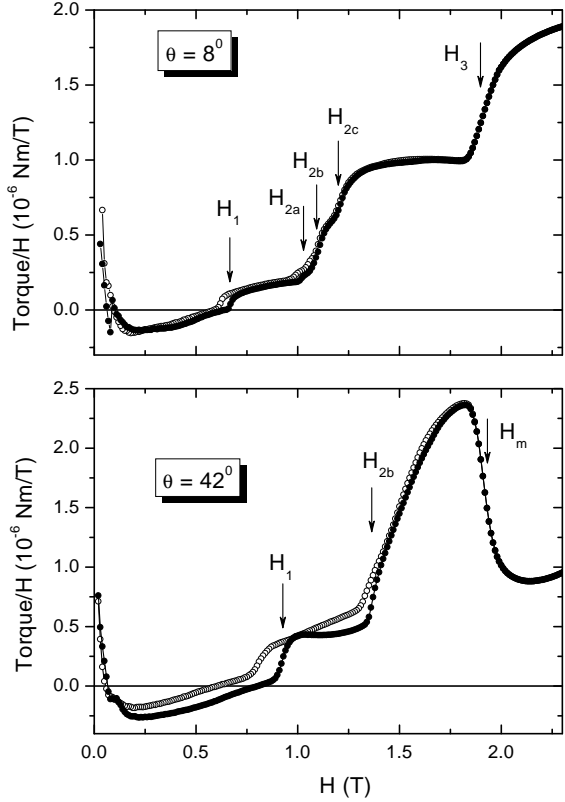}
\caption{Dependences $\tau(H)/H$ for increasing and decreasing
magnetic field (filled and empty circles, respectively) for field
directions close to ($\theta \approx 8^{\circ}$) and far from
($\theta \approx 42^{\circ}$) an easy axis. Positions of the
transition fields are shown by arrows.}
\end{figure}

\begin{figure}[t]
\includegraphics[width=0.86\linewidth]{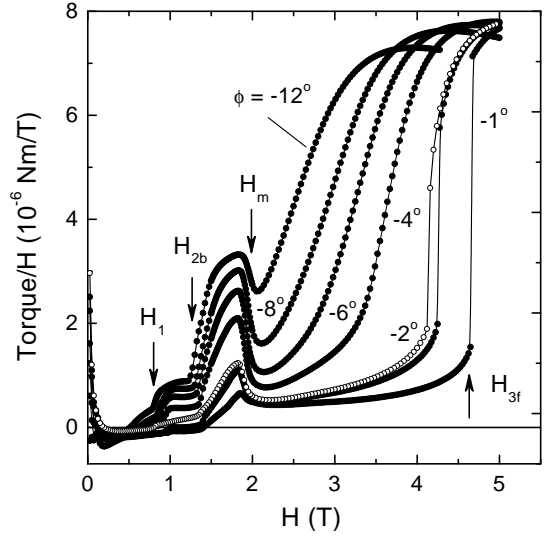}
\caption{$\tau(H)/H$ dependences (increasing field) for different
angles $\phi$ relative to the closest hard axis $\langle 110
\rangle$. Positions of the critical fields $H_{1}$, $H_{2b}$,
$H_{m}$ and $H_{3f}$ are shown by arrows for some of the curves.
For $\phi=-2^{\circ}$, the curves for increasing and decreasing
(empty circles) field are shown. A dramatic increase in $H_{3f}$
when field direction goes to
$|\phi|=0$ is evident. } 
\end{figure}

\begin{figure}[t]
\includegraphics[width=0.8\linewidth]{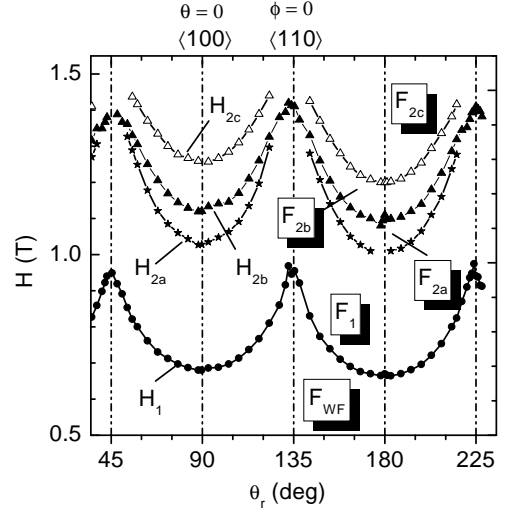}
\vspace{-6pt} \caption{Low-field part of the angular phase diagram
of metamagnetic states in ErNi$_{2}$B$_2$C at $T=1.9$~K.
$\theta_{r}$ is angular position of the sample rotator. The
positions of easy, $\langle 100 \rangle$, and hard, $\langle 110
\rangle$, axes in one of the quadrants are shown. The $H_{1}$,
$H_{2a}$, $H_{2b}$ and $H_{2c}$ are critical fields of
metamagnetic transitions. The symbols $F_{WF}$, F$_1$, F$_{2a}$,
F$_{2b}$, F$_{2c}$ show regions of existence for different phases.
 }
\end{figure}

\par
Results of the neutron \cite{camp,jensen} and magnetization
\cite{canf2,budko} studies  at $T=2$~K are mutually consistent
only for the first two metamagnetic transitions. Detailed neutron
studies \cite{jensen} give a rather complex picture of metamagnetic
states. In the easy direction, five metamagnetic transitions (with
considerable changes in magnetic structure) were found, but only
four SDW magnetic structures were distinguished. For the hard
direction, three SDW phases were seen, but for a considerable
field range the metamagnetic states are mixed-phase states.
For both directions a two-domain magnetic  state was found with
modulation vectors in the domains being perpendicular to each
other.
\par
In this study, a PPMS Model 550 Torque Magnetometer  is used to
study angular dependence of the metamagnetic transitions. It measures
the torque $\vec{\tau} = \mathbf{M}\times\mathbf{H}$, so that
$\tau = MH \sin(\beta)$, where $\beta$ is the angle between the
external magnetic field and the magnetization.  A small (about 0.08
mg) single-crystal rectangular plate of ErNi$_2$B$_2$C
 was used.
The torque was measured as a function of magnetic field for
different constant angles, or varying angular direction of the
field for different constant magnetic fields.
\par
The metamagnetic transitions manifest themselves as sharp changes
in field dependences $\tau(H)/H$ at critical fields as shown in
Figs. 1 and 2. The angular phase diagram for ErNi$_2$B$_2$C
obtained is shown in Figs. 3 and 4. The critical fields were
defined as the fields of the inflection point of the $\tau(H)/H$
curves, recorded for increasing field. These points were found
using derivatives. Some points in the phase diagram were
determined from torque angular dependences recorded for different
fixed fields (Fig. 4).
\par
We have found that the sequence of transitions depends crucially on
the angular position of the magnetic field. In the range $-30^{\circ}
\lesssim \theta \lesssim 30^{\circ}$, with increasing field
transitions at critical fields, denoted as $H_{1}$, $H_{2a}$,
$H_{2b}$, $H_{2c}$ and $H_{3}$ take place (see $\tau(H)/H$ curve
for $\theta\approx 8^{\circ}$ in Fig.~1 and angular diagram in
Figs. 3 and 4). In the adjoining angular region, $-15^{\circ}
\lesssim \phi \lesssim 15^{\circ}$, the sequence of transitions
includes critical fields $H_{1}$, $H_{2b}$, $H_{m}$ and $H_{3f}$
(Figs. 1-4). At fields $H_{3}$ and $H_{3f}$ transitions to the
saturated paramagnetic state take place. The notation for
metamagnetic phases is clear from Figs. 3 and 4. Phases $F_{WF}$
and $F_{PM}$ are the initial WF and the final saturated paramagnetic
states, respectively.
\par
In this study, new metamagnetic states have been found. In the
range 1.0 T $\leq H \leq$ 1.4 T  three closely-spaced transitions
(at $H=H_{2a}$, $H_{2b}$ and $H_{2c}$) for $|\theta|\lesssim
30^{\circ}$ and only one (at $H=H_{2b}$) for $30^{\circ} \lesssim
|\theta| \lesssim 45^{\circ}$ ($|\phi| \lesssim 15^{\circ}$) were
seen (Figs. 1--4).  These critical fields are proportional to
$1/\cos(\theta)$ with $H_{2a}(\theta=0)\approx 1.02$~T,
$H_{2b}(\theta=0)\approx 1.13$~T and $H_{2c}(\theta=0)\approx
1.27$~T.  In a previous longitudinal
magnetization study only two transitions (at
fields which correspond to $H_{2b}$ and $H_{2c}$) were found in an
easy direction at $T=2$~K within this field range and one (at
$H_{2b}$) for the field in a hard direction \cite{budko}.
Only one (that at $H=1.15T$) of
the magnetostructural transitions (at $T=1.8$~K in an easy direction above $H=1$~T)
found in the neutron study \cite{jensen}
corresponds to results of
the longitudinal magnetization \cite{canf1,budko} and this torque
study. The rather weak (but quite clear) transition at $H_{2a}$,
found in this study, was not revealed previously in any study. The
high precision and sensitivity
of the torque measurements to the normal component of $M$ permit its observation.

\begin{figure}[h]
\includegraphics [width=0.92\linewidth]{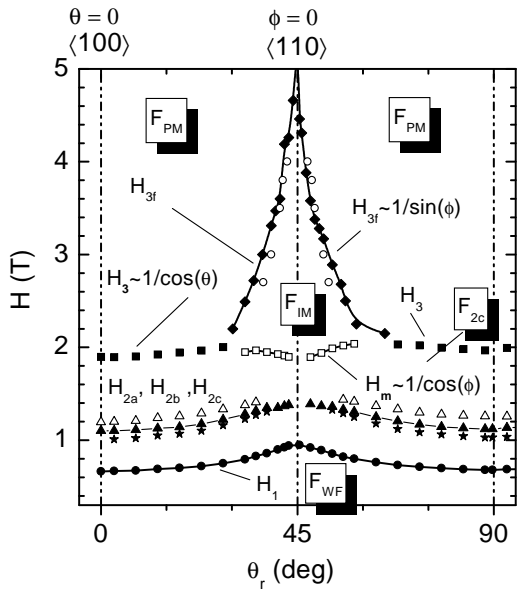}
\vspace{-6pt} \caption{Angular phase diagram of metamagnetic
states in ErNi$_{2}$B$_2$C at $T=1.9$~K for the field range
studied. The angle $\theta_{r}$ is the angle on the sample
rotator. The positions of easy, $\langle 100 \rangle$, and hard,
$\langle 110 \rangle$, axes are shown.  The symbols $F_{WF}$,
F$_{2c}$, $F_{IM}$, and $F_{PM}$ show regions of existence for
different phases.  All data were obtained from $\tau(H)$ curves
for different fixed angles except for the open circles
which were, obtained from torque angular dependences. }
\vspace{-6pt}
\end{figure}

\par
New results were found for fields above 2 T (Fig. 4): a
final transition to the saturated paramagnetic state proceeds
either directly from phase $F_{2c}$ ($F_{2c}\rightarrow F_{PM}$)
for $|\theta|\lesssim 30^{\circ}$ or through an intermediate phase
$F_{IM}$ ($F_{2c}\rightarrow F_{IM}\rightarrow F_{PM}$) for field
directions close enough to a hard axis ($30^{\circ} \lesssim
|\theta| \lesssim 45^{\circ}$). The critical field $H_3$ for the
direct transition ($F_{2c}\rightarrow F_{PM}$) follows
approximately the angular relation
$H_{3}(\theta)=H_{3}(0)/\cos(\theta)$ with $H_{3}(0)\approx
1.93$~T (for $|\theta|\lesssim 10^{\circ}$). Transition to the
intermediate state ($F_{2c}\rightarrow F_{IM}$) takes place at
$H_{m}$ which is approximately described by
$H_{m}(\phi)=H_{m}(0)/\cos(\phi)$ with $H_{m}(0)\approx 1.92$~T.
The critical field $H_{3f}$ for transition to the paramagnetic
state from the intermediate state $F_{IM}$ appears to approach
infinity when the field direction approaches a hard axis
($\phi\rightarrow 0$, which is equivalent to $\theta\rightarrow
45^{\circ}$). Below $H\approx 2.7$~T, the following relation is
found to be roughly obeyed: $H_{3f}(\phi)=H_{3f0}/\sin(\phi)$
with $H_{3f0}\approx 0.62$~T.
\par
All critical fields within the range $|\theta|\lesssim 30^{\circ}$
are found to be proportional to $1/\cos(\theta)$. In this case,
field dependences $\tau(H)/H = M(H)\sin(\beta)$  represent $M(H)$
behavior under and between metamagnetic transitions. They show a
(upper part of Fig.~1) sharp increases in $M(H)$ at the
metamagnetic transitions and more smooth (or nearly
field-independent) behavior between them as found previously
\cite{canf1}  for ErNi$_{2}$B$_2$C for $\theta < 30^{\circ}$ .
\par
When the field direction is close to a hard axis, transition to the
saturated paramagnetic state proceeds through the intermediate
phase $F_{IM}$ (Fig. 4). The transition ($F_{2c}\rightarrow
F_{IM}$) at field $H_m$ is accompanied by a dramatic decrease in
torque (Figs. 1 and 2) which means that the averaged net
magnetization is rotated towards a hard axis. For field direction
in the immediate vicinity of a hard axis, the torque is nearly zero
after completion of the transition. This implies that the net
magnetization is oriented roughly along a hard axis in the phase
$F_{IM}$ for $\phi\rightarrow 0$.
\par
If the Er moments align along an easy axis for any
metamagnetic state as expected, the phase $F_{IM}$ may be
either non-collinear or a mixed state. In either case the net
magnetization can be directed along the hard axis with
$\phi\rightarrow 0$. Since neutron diffraction studies
\cite{camp,jensen} have revealed no indication
of a non-collinear phase, a mixed state appears to be more
likely, since easy axes $\langle 100 \rangle$
and $\langle 010 \rangle$ are equivalent for tetragonal
symmetry.  This is consistent with neutron observation of a
two-domain magnetic structure with perpendicular
modulation vectors (and, accordingly, ordered moments)
equally populated for field direction along a hard
axis \cite{jensen}. Thus, the net magnetization would be
directed along this axis as well, providing a basis not only
for a satisfactory explanation of zero torque magnitude at
$\phi\approx 0$, but also for the origin of the intermediate phase. This
phase is probably a two-domain state where domains
of each type are saturated paramagnetics with different directions
of magnetization. The final transition $F_{IM}\rightarrow F_{PM}$ is
thus a transition from the two-domain to a single-domain state.
This transition was observed \cite{canf1} as a very slight
increase in longitudinal magnetization at a critical field that diverged as
$\phi\rightarrow 0$, but the nature of the transition was not
understood at that time. Results of this torque study make this
quite clear.
\par
This research was supported by the Robert A Welch Foundation
(A-0514), NSF (DMR-0315476 and DMR-0422949).

\end{document}